\documentclass{wicc07}
\usepackage[spanish]{babel}

\usepackage{graphicx}
\usepackage{times}
\usepackage{hyperref}
\hypersetup{
	pdftitle 	={M\'{e}todos para la Selecci\'{o}n y el Ajuste de Caracter\'{i}sticas en el Problema de la Detecci\'{o}n de Spam},	
	pdfauthor 	={Carlos M. Lorenzetti, Roc\'{i}o L. Cecchini, Ana G. Maguitman, Andr\'{a}s A. Bencz\'{u}r}, 	
	pdfsubject 	={Workshop de Investigadores en Ciencias de la Computaci\'{o}n, WICC 2010}
}

\begin{document}

\pagestyle{empty}
\title{\vspace{-1.5cm}\textbf{M\'{e}todos para la Selecci\'{o}n y el Ajuste de Caracter\'{i}sticas en el Problema de la Detecci\'{o}n de Spam}}

\author{
	\textbf{Carlos M. Lorenzetti}$^{\dagger\S}$ \ \ \ 
	\textbf{Roc\'{i}o L. Cecchini}$^{\dagger\ast}$ \ \
	\textbf{Ana G. Maguitman}$^{\dagger\S}$ \ \ \
	\textbf{Andr\'{a}s A. Bencz\'{u}r}$^\ddagger$\vspace{1mm}\\
	\small$^\S$Laboratorio de Inv. y Des. en Inteligencia Artificial\\
	\small$^\ast$Laboratorio de Inv. y Des. en Computaci\'{o}n Cient\'{i}fica\\
	\small$^\dagger$Departamento de Cs. e Ing. de la Computaci\'{o}n -- Universidad Nacional del Sur\\
	\small\tt \{cml,rlc,agm\}@cs.uns.edu.ar \\
	\small$^\ddagger$Data Mining and Web search Research Group -- Informatics Laboratory\\
	\small Computer and Automation Research Institute -- Hungarian Academy of Sciences\\
	\small\tt benczur@ilab.sztaki.hu
}

\maketitle
\thispagestyle{empty}

\section{Introducci\'{o}n}

El correo electr\'{o}nico es quiz\'{a}s la aplicaci\'{o}n que m\'{a}s tr\'{a}fico genera en la Internet. Es utilizado por millones de personas para comunicarse alrededor del mundo y es una aplicaci\'{o}n de misi\'{o}n cr\'{i}tica para muchos negocios. En la \'{u}ltima d\'{e}cada la avalancha de correo no deseado (Spam) ha sido el mayor problema para los usuarios del correo electr\'{o}nico, ya que diariamente una cantidad arrolladora de spam entra en las bandejas de los usuarios. En 2004, se estim\'{o} que el 62\% de todos los correos que se generaron fueron spam~\cite{ion00naive}. El spam no solo es frustrante para muchos usuarios, sino que tambi\'{e}n compromete a la infraestructura tecnol\'{o}gica de las empresas, costando dinero a causa de la p\'{e}rdida de productividad. En los \'{u}ltimos a\~nos, el spam ha evolucionado desde ser una molestia a ser un serio riesgo en la seguridad, llegando a ser el principal medio para el robo de informaci\'{o}n personal, as\'{i} como tambi\'{e}n para la proliferaci\'{o}n de software malicioso.

Muchas alternativas se han propuesto para solucionar el problema, desde protocolos de autenticaci\'{o}n del remitente a, incluso, cobrarles 
dinero a los remitentes \cite{goodman05stopping}. Otra alternativa prometedora es el uso de filtros basados en contenido capaces de discriminar autom\'{a}ticamente entre mensajes spam y mensajes leg\'{i}timos. Los m\'{e}todos de Aprendizaje Automatizado son atractivos para realizar esta tarea ya que son capaces de adaptarse a las caracter\'{i}sticas evolutivas del spam, cont\'{a}ndose adem\'{a}s con disponibilidad de datos para entrenar tales modelos. Sin embargo, uno de los aspectos m\'{a}s frustrantes del spam es que cambia continuamente para adaptarse a las nuevas t\'{e}cnicas que intentan detenerlo. Cada vez que se lo ataca de alguna manera, los generadores de spam encuentran una manera de eludir este ataque. Esta carrera ha llevado a una coevoluci\'{o}n continua y a un aumento del nivel de sofisticaci\'{o}n de ambas partes~\cite{goodman05stopping}. Otra diferencia con respecto a muchas tareas en la clasificaci\'{o}n de texto consiste en que el costo de un error en la clasificaci\'{o}n est\'{a} fuertemente sesgado: etiquetar un correo leg\'{i}timo como spam, usualmente llamado \textit{falso positivo}, trae peores consecuencias que el caso inverso.

La detecci\'{o}n de spam web puede verse como un problema de clasificaci\'{o}n. Para detectar p\'{a}ginas web spam, construimos un clasificador para etiquetar una dada p\'{a}gina como spam o como no spam. Centr\'{a}ndonos en el an\'{a}lisis del contenido sem\'{a}ntico de los correos y de las p\'{a}ginas, se han estudiado varias t\'{e}cnicas de clasificaci\'{o}n de texto basadas en m\'{e}todos de Aprendizaje Automatizado y Reconocimiento de Patrones, debido principalmente a su mayor capacidad de generalizaci\'{o}n. Las t\'{e}cnicas de clasificaci\'{o}n de texto (ver \cite{sebastiani02machine}, para una revisi\'{o}n detallada) se aplican b\'{a}sicamente a documentos de texto representados en formato ASCII no estructurado, en formatos estructurados como HTML y tambi\'{e}n se aplican a mensajes de correo electr\'{o}nico.

El proceso de clasificaci\'{o}n comienza en la fase de entrenamiento y necesita representar el texto plano que contienen los documentos, por esto el primer paso transforma los documentos a alguna representaci\'{o}n interna. Luego se construye un \textit{vocabulario} con todos los t\'{e}rminos que se encontraron en los documentos, para luego pasar a una fase de extracci\'{o}n de caracter\'{i}sticas en donde, por lo general, se reduce la cardinalidad de las mismas. Esto se lleva a cabo mediante la eliminaci\'{o}n de signos de puntuaci\'{o}n y de palabras muy frecuentes, y por el stemming (reducci\'{o}n de las palabras a su palabra ra\'{i}z o stem), con el prop\'{o}sito de descartar t\'{e}rminos no discriminantes y de reducir el tama\~no del vocabulario (y por lo tanto, de la complejidad computacional). Finalmente se representa el documento como un vector de longitud fija de caracter\'{i}sticas, en el cual cada componente (usualmente un n\'{u}mero real) est\'{a} asociado a un t\'{e}rmino del vocabulario. Los t\'{e}rminos usualmente corresponden a palabras individuales, o a frases que se encuentran en los documentos de entrenamiento. Las t\'{e}cnicas de extracci\'{o}n de caracter\'{i}sticas m\'{a}s simples est\'{a}n basadas en un m\'{e}todo de \textit{bolsa de palabras}, en donde solo se tienen en cuenta la ocurrencia de los t\'{e}rminos y se descarta la informaci\'{o}n de su posici\'{o}n dentro del documento. Las caracter\'{i}sticas m\'{a}s comunes son la ocurrencia de la palabra (valor booleano), el n\'{u}mero de ocurrencias (valor entero), o su frecuencia relativa a la longitud del documentos (valor real). Una caracter\'{i}stica llamada TFIDF tiene en cuenta el n\'{u}mero de apariciones en el documento y en todos los documentos de entrenamiento.

Los clasificadores estad\'{i}sticos pueden aplicarse a la representaci\'{o}n vectorial de caracter\'{i}sticas. Las principales t\'{e}cnicas analizadas hasta hoy en este contexto para el filtrado de spam est\'{a}n basadas en el clasificador de texto Bayes Na\"{i}ve \cite{mccallum98comparison} y en los llamados \textquotedblleft filtros Bayesianos'' \cite{sahami98bayesian, graham02plan}. Dado su rendimiento en tareas de clasificaci\'{o}n de texto, tambi\'{e}n se ha investigado el uso de clasificadores M\'{a}quina de Vectores de Soporte (SVM, Support Vector Machine~\cite{drucker99svm,zhang04evaluation}).

\section{L\'{i}nea de Investigaci\'{o}n Propuesta}\label{sec:linea}
\begin{figure*}[!ht]
\centerline{\includegraphics[width=160mm]{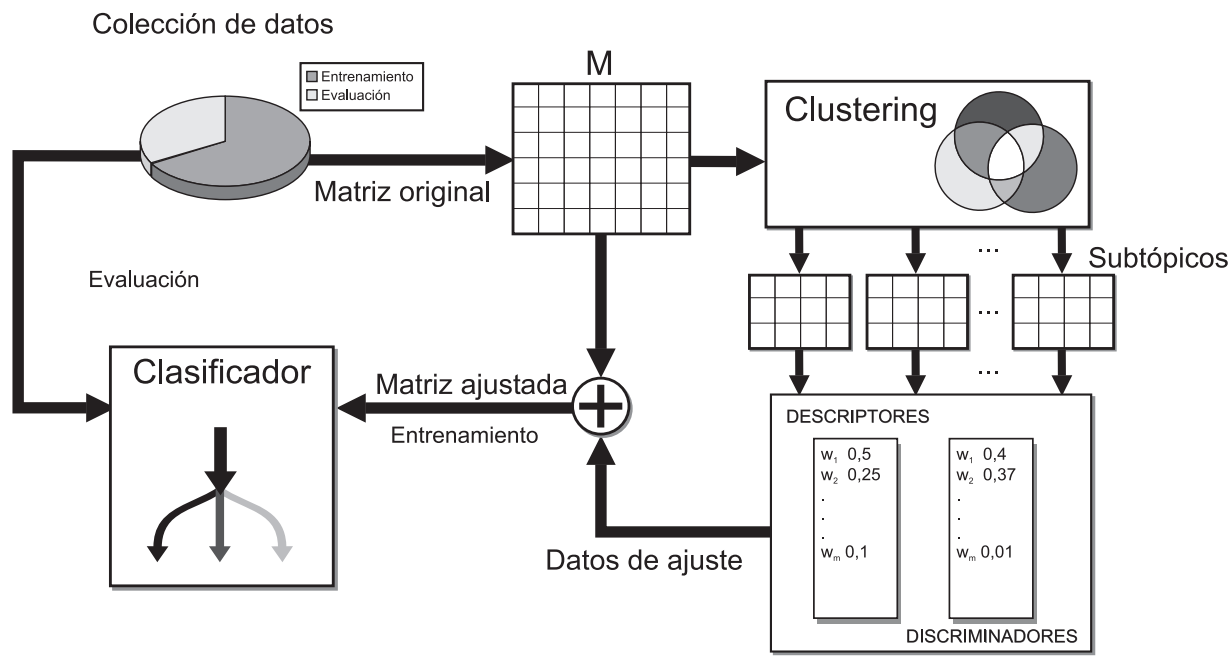}}
\caption{Diagrama esquem\'{a}tico de la propuesta}
\label{fig:diagram}
\end{figure*}

Como se dijo en la secci\'{o}n previa, la identificaci\'{o}n de spam puede verse como un problema de clasificaci\'{o}n. Por lo tanto proponemos un algoritmo que utiliza un clasificador como uno de sus componentes. Nuestra propuesta no incluye el desarrollo de un clasificador en s\'{i} mismo, sino que plantea un ajuste en los datos de entrada del conjunto de entrenamiento del clasificador con el objetivo de mejorar su rendimiento. El esquema general del sistema se muestra en la figura~\ref{fig:diagram} y se describe a continuaci\'{o}n.
\subsection{Clustering}
Dada la heterogeneidad que posee el spam, no puede asumirse que todo spam se asocia a un \'{u}nico t\'{o}pico. Es por esto que proponemos, como primera etapa, la utilizaci\'{o}n de un algoritmo de clustering que dividir\'{a} a los documentos en subt\'{o}picos m\'{a}s peque\~nos esperando con esto una mejora en el rendimiento global del algoritmo. Una lista detallada de los algoritmos disponibles para este prop\'{o}sito puede encontrarse en~\cite{berkhin06clustsurv}.

\subsection{Descriptores y Discriminadores}
Una vez que los datos de entrada se encuentran agrupados en subt\'{o}picos m\'{a}s espec\'{i}ficos, tomamos cada uno de ellos y calculamos los pesos de los t\'{e}rminos en ellos como descriptores y discriminadores de estos subt\'{o}picos. 

En~\cite{maguitman04dynamic} proponemos estudiar el poder descriptivo y discriminante de un t\'{e}rmino en base a su distribuci\'{o}n a trav\'{e}s de los t\'{o}picos de las p\'{a}ginas recuperadas por un motor de b\'{u}squeda.

Para distinguir entre descriptores y discriminadores de t\'{o}picos argumentamos que \textit{buenos descriptores de t\'{o}picos} pueden encontrarse buscando aquellos t\'{e}rminos que aparecen \underline{con frecuencia} en documentos relacionados con el t\'{o}pico deseado. Por otro lado, \textit{buenos discriminadores de t\'{o}picos} pueden hallarse buscando t\'{e}rminos que aparecen \underline{solo} en documentos relacionados con el t\'{o}pico deseado. Ambos tipos de t\'{e}rminos son importantes a la hora de generar consultas. Utilizar t\'{e}rminos descriptores del t\'{o}pico mejora el problema de los resultados falso negativo porque aparecen frecuentemente en p\'{a}ginas relevantes. De la misma manera, los buenos discriminadores de t\'{o}picos ayudan a reducir el problema de los falsos positivos, ya que aparecen principalmente en p\'{a}ginas relevantes.

Esta etapa da como resultado listas de t\'{e}rminos con informaci\'{o}n asociada a la importancia de los mismos como descriptores y discriminadores. Dicha informaci\'{o}n se utilizar\'{a} para ajustar la matriz de datos de entrenamiento para reflejar de forma m\'{a}s fidedigna los pesos de los t\'{e}rminos en los documentos.
\subsection{Clasificador}\label{subsec:clasif}
Los clasificadores son implementados a partir de un conjunto de instancias o ejemplos previamente etiquetados, en donde cada ejemplo tiene un vector de atributos o caracter\'{i}sticas. En general, en los conjuntos de datos que utilizaremos en nuestras evaluaciones (descriptos en la secci\'{o}n~\ref{sec:datos}) las etiquetas fueron determinadas por personas.

La clasificaci\'{o}n involucra la creaci\'{o}n de un modelo durante la etapa de entrenamiento que predecir\'{a} la etiqueta de cada instancia del conjunto de testeo usando los valores del vector de caracter\'{i}sticas. Para construir el clasificador, primero lo entrenamos sobre un n\'{u}mero de ejemplos del conjunto etiquetado de entrenamiento y determinamos los par\'{a}metros de nuestro clasificador. Durante la etapa de testeo, el clasificador examina el vector de caracter\'{i}sticas de forma conjunta para determinar si una p\'{a}gina pertenece a una dada categor\'{i}a o no. La evaluaci\'{o}n del clasificador se realiza en la etapa de testeo comparando, para cada instancia, la etiqueta calculada por el clasificador con la asignada a esa instancia. 

Una lista detallada de los algoritmos disponibles para este prop\'{o}sito puede encontrarse en~\cite{jain99data, qi09webpage}. Se prev\'{e} utilizar el entorno Weka~\cite{hall09weka_upd} en esta etapa.

\section{Evaluaci\'{o}n}\label{sec:datos}
Para la evaluaci\'{o}n de nuestra propuesta utilizaremos distintos conjuntos de datos disponibles, como por ejemplo el conjunto de datos UK-2007 del workshop internacional AirWeb~\cite{airweb09}, el conjunto de datos de la conferencia internacional ECML PKDD~\cite{hotho08challenge}, el conjunto de datos del track de Spam de la conferencia TREC~\cite{cormack07spamtrec} y el corpus de correos electr\'{o}nicos SpamAssassin~\cite{spamassassin}.
Para analizar la eficacia del m\'{e}todo propuesto evaluaremos el rendimiento del clasificador. Para ello utilizaremos las m\'{e}tricas est\'{a}ndares de evaluaci\'{o}n, como precisi\'{o}n, cobertura, F-score, Media Geom\'{e}trica, \'{a}rea bajo la curva ROC, \'{a}rea bajo la curva Precisi\'{o}n-Cobertura y estad\'{i}sticas de Kolmogorov-Smirnov.

\section{Conclusiones}\label{sec:conclusions}

La t\'{e}cnica propuesta en este trabajo ataca uno de los problemas m\'{a}s grandes a los que se deben enfrentar los usuarios de los sistemas de informaci\'{o}n actuales. Mejorar la representaci\'{o}n de los documentos mediante el uso de vocabularios m\'{a}s representativos, as\'{i} como el ajuste de los datos realizado a trav\'{e}s de la detecci\'{o}n de buenos descriptores y discriminadores ha mostrado ser efectivo en otras \'{a}reas de recuperaci\'{o}n de informaci\'{o}n~\cite{Lorenzetti2009semisupervised, lorenzetti08tuning}. Anticipamos  que aplicar estos m\'{e}todos ser\'{a} ventajoso para abordar diversos  problemas de clasificaci\'{o}n, en particular en el \'{a}mbito de la detecci\'{o}n de spam.

Nuestro trabajo est\'{a} relacionado con muchos estudios previos sobre clasificaci\'{o}n de p\'{a}ginas web spam basados en caracter\'{i}sticas. Desde los comienzos de la World Wide Web ha existido una necesidad de calificar a las p\'{a}ginas de acuerdo a su relevancia con una dada consulta. Sin embargo se ha puesto un nuevo \'{e}nfasis al problema dadas las grandes ganancias que genera la publicidad a trav\'{e}s de Internet. La clasificaci\'{o}n de spam web es uno de los desaf\'{i}os m\'{a}s importantes de los motores de b\'{u}squeda~\cite{henzinger03challenges}, en particular debido a la degradaci\'{o}n de la calidad de sus resultados. Un m\'{e}todo prometedor para la identificaci\'{o}n del spam web es la utilizaci\'{o}n de la informaci\'{o}n de los enlaces web que contienen las p\'{a}ginas~\cite{davison00recognizing,amitay03connectivity,becchetti2006using, gyongyi05alliances, wu05identifying}. Por otro lado recientemente se ha estudiado la clasificaci\'{o}n de spam web bas\'{a}ndose en el contenido de la p\'{a}gina~\cite{ntoulas06detecting,fetterly04spam}. La detecci\'{o}n de spam en blogs se estudi\'{o} en~\cite{mishne2005bbs}.

Todos estos ejemplos son solo los primeros pasos en el combate contra el spam: la naturaleza necesariamente adversaria de la tarea conlleva a un problema que evoluciona r\'{a}pidamente, y esta caracter\'{i}stica (de tener que buscar t\'{e}cnicas que sean exitosas a la luz de la adaptaci\'{o}n del enemigo) es algo nuevo en la comunidad de Aprendizaje Automatizado y trae consigo numerosos desaf\'{i}os y oportunidades de investigaci\'{o}n.

\begin{footnotesize}

\end{footnotesize}

\end{document}